\documentclass[10pt]{amsart}

\usepackage{amsmath}
\usepackage{amssymb}
\usepackage{amsthm}

\newtheorem{theorem}{Theorem}[section]
\newtheorem{conjecture}{Conjecture}[section]
\newtheorem{proposition}{Proposition}[section]
\newtheorem{remark}{Remark}[section]

\newcommand{\Z}{{\mathbb Z}}

\newcommand{\aff}{\hbox{Aff}}

\begin{document}

\title{Periodic cellular automata and Bethe ansatz}

\author{Atsuo Kuniba}
\address{Institute of Physics, University of Tokyo, Tokyo 153-8902, Japan}
\email{atsuo@gokutan.c.u-tokyo.ac.jp}

\author{Akira Takenouchi}
\address{Institute of Physics, University of Tokyo, Tokyo 153-8902, Japan}
\email{takenouchi@gokutan.c.u-tokyo.ac.jp}
\date{}

\maketitle

\begin{abstract}
We review and generalize the recent progress in a 
soliton cellular automaton known as the periodic box-ball system.
It has the extended affine Weyl group symmetry and admits 
the commuting transfer matrix method and the Bethe ansatz at $q=0$.
Explicit formulas are proposed for the dynamical period
and the number of states characterized by conserved quantities.
\end{abstract}


\section{Introduction}\label{sec:1}

In \cite{KT}, a class of periodic soliton cellular automata is introduced 
associated with non-exceptional quantum affine algebras. 
The dynamical period and a 
state counting formula are proposed by the 
Bethe ansatz at $q=0$ \cite{KN}.
In this paper we review and generalize 
the results on the $A^{(1)}_n$ case, where 
the associated automaton is known as the 
periodic box-ball system \cite{MIT,YYT}.
The box-ball system was originally introduced 
on the infinite lattice without boundary \cite{TS,T}.
Here is a collision of two solitons with amplitudes 3 and 1 
interchanging internal degrees of freedom with a phase shift:

\noindent
\begin{center}
$\cdots 1114221111131111111111111111\cdots$\\ 
$\cdots 1111114221113111111111111111 \cdots$\\ 
$\cdots 1111111114221311111111111111\cdots$\\
$\cdots 1111111111114232111111111111\cdots$\\
$\cdots 1111111111111121432111111111 \cdots$\\
$\cdots 1111111111111112111432111111 \cdots$\\
$\cdots 1111111111111111211111432111 \cdots$
\end{center}

\noindent
The system was identified \cite{HHIKTT,FOY} 
with a solvable lattice model \cite{B} at $q=0$, which led to 
a direct formulation by the crystal base theory \cite{K}
and generalizations to the soliton cellular automata 
with quantum group symmetry \cite{HKT,HKOTY}.
Here we develop the approach to the periodic case 
in \cite{KT} 
further by combining 
the commuting transfer matrix method \cite{B}
and the Bethe ansatz \cite{Be} at $q=0$.

In section \ref{sec:2} we formulate 
the most general periodic 
automaton for $\ensuremath{\mathfrak{g}_n}= A^{(1)}$ 
in terms of the crystal theory.
A commuting family of time evolutions $\{T^{(a)}_j\}$ is introduced as 
commuting transfer matrices under periodic boundary condition.
The associated conserved quantities 
form an $n$-tuple of Young diagrams $m=(m^{(1)},\ldots, m^{(n)})$,
which we call the {\em soliton content}.

In section \ref{sec:3} we invoke 
the Bethe ansatz at $q=0$ \cite{KN} to study the Bethe eigenvalue 
$\Lambda^{(r)}_l$ relevant to $T^{(r)}_l$.
The Bethe equation is linearized into the string center equation and 
the $\Lambda^{(r)}_l$ is shown to be a root of unity.
We also recall an explicit weight multiplicity formula 
obtained by counting the off-diagonal solutions to the 
string center equation \cite{KN}. 
It is a version of the fermionic formula
called the combinatorial completeness of the 
string hypothesis at $q=0$. 
These results are parameterized with the number of
strings, which we call the {\em string content}.

In section \ref{sec:4} two applications of 
the results in section \ref{sec:3} 
are presented under the identification of the 
soliton and the string contents.
First we relate the root of unity in the Bethe eigenvalue $\Lambda^{(r)}_l$ 
to the dynamical period of 
the periodic $A^{(1)}_n$ automaton under the time evolution $T^{(r)}_l$.
Second we connect each summand 
in the weight multiplicity formula \cite{KN}
to the number of states characterized by 
conserved quantities.

In \cite{KT}, similar results have been announced concerning the 
highest states.
Our approach here is based on conserved quantities and 
covers a wider class of states 
without recourse to the combinatorial Bethe ansatz at 
$q=1$ \cite{KKR}.
We expect parallel results in general 
$\ensuremath{\mathfrak{g}_n}$.
In fact all the essential claims in this paper 
make sense also for 
$\ensuremath{\mathfrak{g}_n}=D^{(1)}_n$ and $E^{(1)}_{6,7,8}$.
Our formulas (\ref{eq:lcmsl2}) and (\ref{omega}) 
include the results in \cite{YYT} 
proved by a different approach
as the case 
$\ensuremath{\mathfrak{g}_n}=A^{(1)}_1$  with 
$B = (B^{1,1})^{\otimes L}$ and $l=\infty$.
For the standard notation and facts in the crystal theory, 
we refer to \cite{K,KMN,HKOTY}.

\section{Periodic $A^{(1)}_n$ automaton}\label{sec:2}
Let $B^{a,j}\;(1\!\le\! a \!\le\! n, j \in \Z_{\ge 1})$ 
be the crystal \cite{KMN} of 
the Kirillov-Reshetikhin module $W^{(a)}_j$ over $U_q(A^{(1)}_n)$.
Elements of $B^{a,j}$ are labeled with semistandard tableaux 
on an $a \times j$ rectangular Young diagram with letters 
$\{1,2,\ldots, n+1\}$.
For example when $n=2$, one has
$B^{1,1}=\{1,2,3\}, B^{1,2} = \{11,12,13,22,23,33\}$, 
$B^{2,2}=\Big\{
\small{\begin{array}{cc}1\!&\!1\\[-1mm]2\!&\!2\end{array}},
\small{\begin{array}{cc}1\!&\!1\\[-1mm]2\!&\!3\end{array}},
\small{\begin{array}{cc}1\!&\!1\\[-1mm]3\!&\!3\end{array}},
\small{\begin{array}{cc}1\!&\!2\\[-1mm]2\!&\!3\end{array}},
\small{\begin{array}{cc}1\!&\!2\\[-1mm]3\!&\!3\end{array}},
\small{\begin{array}{cc}2\!&\!2\\[-1mm]3\!&\!3\end{array}}
\Big\}$ as sets.
$\aff(B^{a,j}) = \{\zeta^db\mid b \in B^{a,j}, d \in \Z\}$ denotes the
affine crystal.
The combinatorial $R$ is the isomorphism of affine crystals
$\aff(B^{a,j})\otimes \aff(B^{b,k}) \overset{\sim}{\rightarrow}
\aff(B^{b,k})\otimes \aff(B^{a,j})$ \cite{S}.
It has the form 
$R(\zeta^db\otimes \zeta^ec) = \zeta^{e+H}{\tilde c} \otimes \zeta^{d-H}{\tilde b}$,
where $H = H(b \otimes c)$ is the energy function.
We normalize it so as to attain the maximum at 
$H(u^{a,j} \otimes u^{b,k}) = 0$, 
where $u^{a,j} \in B^{a,j}$ denotes the classically highest element.
We set 
$B = B^{r_1,l_1}\otimes B^{r_2,l_2}\otimes 
\cdots \otimes B^{r_L,l_L}$ 
and write 
$\aff(B^{r_1,l_1})\otimes \cdots \otimes \aff(B^{r_L,l_L})$ simply as 
$\aff(B)$.
An element of $B$ is called a state.
Given a state $p = b_1\otimes \cdots \otimes b_L \in B$, regard it as 
the element $\zeta^0b_1 \otimes \cdots \otimes \zeta^0 b_L \in \aff(B)$ and 
seek an element $v \in B^{r,l}$ such that 
$\zeta^0 v \otimes p \simeq 
(\zeta^{d_1}b'_1\otimes \cdots \otimes \zeta^{d_L}b'_L) \otimes \zeta^{e} v$ under the 
isomorphism 
$\aff(B^{r,l}) \otimes \aff(B) \simeq \aff(B)\otimes \aff(B^{r,l})$.
If such a $v$ exists and 
$\zeta^{d_1}b'_1\otimes \cdots \otimes \zeta^{d_L}b'_L$ 
is unique even if $v$ is not unique,
we say that $p$ is $(r,l)$-{\em evolvable} and 
write $T^{(r)}_l(p) = b'_1\otimes \cdots \otimes b'_L \in B$ and 
$E^{(r)}_l(p) = e=-d_1-\cdots - d_L$.
Otherwise we say that $p$ is not $(r,l)$-evolvable or $T^{(r)}_l(p)=0$.
We formally set $T^{(r)}_l(0)=0$.
$E^{(r)}_l  \in \Z_{\ge 0}$ holds under this normalization.
Our $A^{(1)}_n$ automaton is a dynamical system on $B \sqcup \{ 0 \}$ 
equipped with the family of time evolutions 
$\{T^{(r)}_l\mid 1 \le r \le n, l \in \Z_{\ge 1}\}$. 
$T^{(r)}_l$ is the $q=0$ analogue of the transfer matrix 
in solvable vertex models.
It is invertible and weight preserving on $B$.
Using the Yang-Baxter equation of the combinatorial $R$, 
one can show (cf. \cite{FOY,HKOTY})
\begin{theorem}\label{th:commute}
The commutativity 
$T^{(a)}_jT^{(b)}_k(p) = T^{(b)}_kT^{(a)}_j(p)$ is valid 
for any $(a,j), (b,k)$ and $p \in B$, where 
the both sides are either in $B$ or $0$.
In the former case
$E^{(a)}_j(T^{(b)}_k(p))=E^{(a)}_j(p)$ and 
$E^{(b)}_k(T^{(a)}_j(p))=E^{(b)}_k(p)$ hold.
\end{theorem}
Thus, $\{E^{(a)}_j\mid 1 \le a \le n, j \in \Z_{\ge 1}\}$ 
is a family of conserved quantities.
\begin{conjecture}\label{conj:convergent}
For any $1\!\le\! a\! \le\! n$ and $p \in B$, 
there exists $i\ge 1$ such that 
$T^{(a)}_k(p)\neq 0$ if and only if $k \ge i$.
The limit $\lim_{k\rightarrow \infty}T^{(a)}_k(p) \in B$ exists and 
$E^{(a)}_i(p) < E^{(a)}_{i+1}(p) < \cdots < E^{(a)}_{j}(p) = E^{(a)}_{j+1}(p)
= \cdots$ holds for some $j \ge i$.
\end{conjecture}
Let $S_0, S_1, \ldots, S_n$ be the Weyl  group operators \cite{K}
and ${\rm pr}$ be the promotion operator \cite{S}
acting on $B$ component-wise.  For instance for $A^{(1)}_3$, 
${\rm pr}\Big(\small{\begin{array}{ccc}2\!&\!2\!&\!3\\[-1mm]
 3\!&\!3\!&\!4 \end{array}}\otimes 1344 \Big) = 
\small{\begin{array}{ccc}1\!&\!3\!&\!3 \\[-1mm] 
4\!&\!4\!&\!4\end{array}} \otimes 1124
\in B^{2,3}\otimes B^{1,4}$.
They act on $B$ as the extended affine Weyl group 
$\widetilde{W}(\ensuremath{\mathfrak{g}_n})
=\widetilde{W}(A^{(1)}_n) 
= \langle {\rm pr}, S_0, S_1, \ldots, S_n \rangle$.
\begin{theorem}\label{th:w}
If $T^{(a)}_j(p)\neq 0$, then for any 
$w \in \widetilde{W}(A^{(1)}_n)$, 
the relations $wT^{(a)}_j(p) = T^{(a)}_j(w(p))$ and 
$E^{(a)}_j(w(p)) = E^{(a)}_j(p)$ are valid.
\end{theorem}

A state $p \in B$ is called {\em evolvable} if 
it is $(a,j)$-evolvable for any $(a,j)$.
In Conjecture \ref{conj:convergent}, 
we expect that the convergent limit $T^{(a)}_\infty$ equals 
a translation in $\widetilde{W}(A^{(1)}_n)$.
Compared with $T_l$ in \cite{KT}, the family 
$\{T^{(a)}_j\}$ here 
is more general and enjoys a larger symmetry  
$\widetilde{W}(A^{(1)}_n)$.
Define the subset of $B$ by
\begin{equation}\label{eq:pm}
P(m) = \{ p \in B \mid p: \hbox{evolvable}, 
E^{(a)}_j(p) = \sum_{k\ge 1}\min(j,k)m^{(a)}_k\}.
\end{equation}
Pictorially, $m=(m^{(1)}, \ldots, m^{(n)})$ is the 
$n$-tuple of Young diagrams and 
$E^{(a)}_k$ (resp. $m^{(a)}_k$) is the number of 
nodes in the first $k$ columns 
(resp. number of length $k$ rows) in $m^{(a)}$.
We call $m$ the {\em soliton content}.
\begin{remark}\label{rem:inv}
$P(m)$ is $\widetilde{W}(A^{(1)}_n)$-invariant
due to Theorem $\ref{th:w}$.
\end{remark}

Given $p \in P(m)$, $T^{(a)}_j(p) \in P(m)$ is not necessarily valid.
For instance 
$p = 112233 \in P(((22),(2))) \subset B=(B^{1,1})^{\otimes 6}$ 
but $T^{(2)}_1(p) = 213213$ is not evolvable since $(T^{(2)}_1)^2(p)=0$.
On the other hand one can show
\begin{proposition}\label{prop:exist}
If $p \in P(m)$ and $(T^{(a)}_j)^t(p)\neq 0$ for any $t$, 
then $(T^{(a)}_j)^t(p) \in P(m)$ for any $t$. 
\end{proposition}
Let $W, \Lambda_a, \alpha_a$ be the Weyl group, 
the fundamental weights and the simple roots 
of $A_n$, respectively.
We specify $p^{(a)}_j=p^{(a)}_j(m)$ by (\ref{eq:vacancy}) and set 
\begin{align}\label{lamH}
\lambda(m)=\sum_{a=1}^np^{(a)}_\infty\Lambda_a,\quad
H=\{(a,j)\mid 1 \le a \le n, j\in \Z_{\ge 1}, 
m^{(a)}_j>0\}.
\end{align}
\begin{conjecture}\label{conj:adm}
$P(m) \neq \emptyset$ if and only if $p^{(a)}_j \ge 0$
for all $(a,j) \in H$.
$\{{\rm wt}p \mid p \in P(m)\} = W\lambda(m)$.
\end{conjecture}
The claim on the weights is consistent with the 
$\widetilde{W}(A^{(1)}_n)$-invariance of $P(m)$.

Here is an example of time evolutions 
in $B = B^{1,1}\otimes B^{1,1} \otimes B^{1,3} \otimes
 B^{1,1} \otimes B^{1,1} \otimes B^{1,1} \otimes B^{1,2}$ with 
$\widetilde{W}(A^{(1)}_3)$ symmetry.
The leftmost column is 
$p_0, T^{(1)}_2(p_0), T^{(2)}_1T^{(1)}_2(p_0)$ and 
$T^{(3)}_2T^{(2)}_1T^{(1)}_2(p_0)$
{}from the top to the bottom.
At each time step, the states connected by 
the Weyl group actions $S_0$ and $S_1$ are shown, 
forming commutative diagrams. ($\cdot$ signifies $\otimes$.)
All these states belong to $P(((3111),(21),(1)))$.

\begin{center}
$2 \cdot 1 \cdot 233 \cdot 4 \cdot 1 \cdot 2 \cdot 12
\;\stackrel{S_0}{\mapsto}\;
 2 \cdot 4 \cdot 233 \cdot 4 \cdot 1 \cdot 2 \cdot 24
\;\stackrel{S_1}{\mapsto}\;
 1 \cdot 4 \cdot 133 \cdot 4 \cdot 1 \cdot 2 \cdot 14$\\

$1 \cdot 2 \cdot 123 \cdot 3 \cdot 4 \cdot 1 \cdot 22
\;\;\phantom{\stackrel{S_2}{\mapsto}}\;\;
 4 \cdot 2 \cdot 234 \cdot 3 \cdot 4 \cdot 1 \cdot 22 
\;\;\phantom{\stackrel{S_2}{\mapsto}}\;\;
 4 \cdot 1 \cdot 134 \cdot 3 \cdot 4 \cdot 1 \cdot 12 $\\

$1 \cdot 2 \cdot 112 \cdot 3 \cdot 2 \cdot 3 \cdot 24
\;\;\phantom{\stackrel{S_2}{\mapsto}}\;\;
1 \cdot 2 \cdot 244 \cdot 3 \cdot 2 \cdot 3 \cdot 24
\;\;\phantom{\stackrel{S_2}{\mapsto}}\;\;
1 \cdot 2 \cdot 144 \cdot 3 \cdot 1 \cdot 3 \cdot 14$\\

$2 \cdot 3 \cdot 112 \cdot 4 \cdot 2 \cdot 3 \cdot 12
\;\;\phantom{\stackrel{S_2}{\mapsto}}\;\;
2 \cdot 3 \cdot 244 \cdot 4 \cdot 2 \cdot 3 \cdot 12
\;\;\phantom{\stackrel{S_2}{\mapsto}}\;\;
2 \cdot 3 \cdot 144 \cdot 4 \cdot 1 \cdot 3 \cdot 11$
\end{center}
\begin{remark}
Let $R_j (1\!\le\! j\!\le\!L\!-\! 1)$ 
be the combinatorial $R$ that exchanges the $j$-th and $(j\!+\! 1)$-st 
components in 
$B = B^{r_1,l_1}\otimes \cdots \otimes B^{r_L,l_L}$ 
and $\pi(b_1\otimes \cdots \otimes b_L) 
= b_L\otimes b_1\otimes \cdots \otimes b_{L-1}$.
Together with $R_0 = \pi^{-1}R_1\pi$, they act on 
$\cup_{s \in {\mathfrak S}_L} B^{r_{s_1},l_{s_1}}
\otimes \cdots \otimes B^{r_{s_L},l_{s_L}}$
as the extended affine Weyl group 
$\widetilde{W}(A^{(1)}_{L-1}) = 
\langle \pi, R_0, \ldots, R_{L-1}\rangle$.
Theorem \ref{th:w} is actually valid for 
$w\in \widetilde{W}(\ensuremath{\mathfrak{g}_n}\!=\!A^{(1)}_{n}) 
\times \widetilde{W}(A^{(1)}_{L-1})$ 
as in \cite{KNY}.
In the homogeneous case 
$(r_1,l_1)=\cdots = (r_L,l_L)$, the $\widetilde{W}(A^{(1)}_{L-1})$ 
symmetry shrinks down to the $\pi$-symmetry, 
which is the origin of the adjective ``periodic".  
\end{remark}

\section{Bethe ansatz at $q=0$}\label{sec:3}

Eigenvalues of row transfer matrices in 
trigonometric vertex models are 
given by the analytic Bethe ansatz \cite{R,KS}.
Let
$Q_r(u)= \prod_k\sinh\pi(u-\sqrt{-1}u^{(r)}_k)$ be 
Baxter's $Q$-function where 
$\{u^{(a)}_j\}$ satisfy 
the Bethe equation eq.(2.1) in \cite{KN}.
We set $q = e^{-2\pi\hbar}$ and $\zeta = e^{2\pi u}$.
For the string solution 
(\cite{KN} Definition 2.3), 
the relevant quantity to our $T^{(r)}_l$ is 
the top term of the eigenvalue $\Lambda^{(r)}_l(u)$ 
(cf.\cite{KS} (2.12)):
\begin{equation}\label{la0}
\lim_{q \rightarrow 0}
\frac{Q_r(u-l\hbar)}{Q_r(u+l\hbar)}= 
\zeta^{-E^{(r)}_l}\Lambda^{(r)}_l,
\; 
\Lambda^{(r)}_l:=\prod_{j\alpha}(-z^{(r)}_{j\alpha})^{\min(j,l)}.
\end{equation}
Here $z^{(a)}_{j\alpha}$ 
is the center of the $\alpha$-th string 
having color $a$ and length $j$.
Denote by $m^{(a)}_j$ the number of such strings.
We call the data $m=(m^{(a)}_j)$ the {\em string content}.
The product in (\ref{la0}) is taken over 
$j\in \Z_{\ge 1}$ and $1 \le \alpha \le m^{(r)}_j$.
$E^{(r)}_l$ is given by the same expression as in
(\ref{eq:pm}) as the function of $m$.
At $q=0$ the Bethe equation becomes 
the string center equation (\cite{KN} (2.36)):
\begin{align}
&\prod_{(b,k)\in H} \prod_{\beta = 1}^{m^{(b)}_k}
(z^{(b)}_{k\beta})^{A_{aj\alpha,bk\beta}} =
 (-1)^{p^{(a)}_j+m^{(a)}_j+1},\label{sce0}\\
A_{aj\alpha,bk\beta} &= 
\delta_{ab}\delta_{j k}\delta_{\alpha \beta}
(p^{(a)}_j+m^{(a)}_j) +
C_{ab}\min(j,k) - \delta_{ab}\delta_{j k},\\
p^{(a)}_j &=\sum_{i=1}^L\min(j,l_i)\delta_{a r_i}
- \sum_{(b,k) \in H}C_{ab}\min(j,k)m^{(b)}_k,\label{eq:vacancy}
\end{align}
where $(C_{ab})_{1 \le a,b \le n}$ 
is the Cartan matrix of $A_n$.
To avoid a notational complexity 
we temporally abbreviate the triple indices $a j \alpha$ 
to $j$, $b k \beta$ to $k$ and accordingly $z^{(b)}_{k\beta}$ 
to $z_k$ etc.
Then (\ref{la0}) and (\ref{sce0}) read
\begin{equation}\label{la}
\Lambda^{(r)}_l = \prod_{k}(-z_k)^{\rho_{k}},\quad 
\prod_k (-z_k)^{A_{j,k}} = (-1)^{s_j},
\end{equation}
where $\rho_k$ is given by $\rho_{k} = \delta_{b r}\min(k,l)$
for $k$ corresponding to $bk\beta$, and $s_j$ is an integer.
Note that $A_{j,k}=A_{k,j}$.
Suppose that the $q=0$ eigenvalue satisfies 
$(\Lambda^{(r)}_l)^{{\mathcal P}^{(r)}_l} = \pm 1$ 
for generic solutions to 
the string center equation
\footnote{${\mathcal P}^{(r)}_l$ here should not be confused 
with the symbol in (\ref{eq:vacancy}).}.
It means that there exist integers $\xi_j$ such that 
$\sum_j\xi_jA_{j,k} = {\mathcal P}^{(r)}_l\rho_{k}$, or equivalently
$\xi_j = {\mathcal P}^{(r)}_l\frac{\det A[j]}{\det A}$,
where $A[j]$ denotes the matrix $A=(A_{j,k})$ 
with its $j$-th column replaced by 
$^t(\rho_{1},\rho_{2},\ldots)$.
In view of the condition $\forall \xi_j \in \Z$, 
the minimum integer allowed for ${\mathcal P}^{(r)}_l$ is
$
{\mathcal P}^{(r)}_l = {\rm LCM}\Bigl(1, \, 
\bigcup_k{}^\prime{}\frac{\det A}{\det A[k]} \Bigr)
$, 
where LCM stands for the least common multiple 
and $\cup_k^\prime$ means the union over those $k$ such that 
$A[k] \neq 0$.
Back in the original indices, 
the determinants here can be 
simplified (cf. \cite{KN} (3.9)) 
to those of matrices indexed with $H$:
\begin{equation}\label{lcm}
{\mathcal P}^{(r)}_l= {\rm LCM}\Bigl(1, 
\bigcup_{(b,k) \in H}\!\!\!\!\!{}^\prime\;\;\frac{\det F}{\det F[b,k]} \Bigr),
\end{equation}
where the matrix $F=(F_{aj, bk})_{(a,j), (b,k) \in H}$ is defined by
\begin{equation}\label{F}
F_{aj, bk} = \delta_{ab}\delta_{jk}p^{(a)}_j + C_{a b}\min(j,k)m^{(b)}_k.
\end{equation}
The matrix $F[b,k]$ is obtained {}from $F$ by replacing its 
$(b,k)$-th column as
\begin{equation}
F[b,k]_{aj,cm} = \begin{cases}
F_{aj,cm} & (c,m) \neq (b,k),\\
\delta_{a r}\min(j,l) & (c,m)=(b,k).
\end{cases}
\end{equation}
The union in (\ref{lcm}) is taken over those $(b,k)$ such that 
$\det F[b,k] \neq 0$.

The LCM (\ref{lcm}) can further be simplified 
for $A^{(1)}_1, r=1, \forall r_i = \forall l_i = 1$.
We write $p^{(1)}_j$ just as $p_j$ and parameterize the set
$H = \{ j \in \Z_{\ge 1} \mid m^{(1)}_j > 0\}$ as  
$H = \{(0 < )J_1 < \cdots < J_s \}$. 
Setting $i_k = \min(J_k,l)$ and $i_0 = 0$, one has
\begin{equation}\label{eq:lcmsl2}
{\mathcal P}^{(1)}_l 
= {\rm LCM}\left(1, \bigcup_{k=0}^{t}\!{}^\prime
\frac{p_{i_{k+1}}p_{i_k}}
{(i_{k+1}-i_k)p_{i_s}}\right),
\end{equation}
where $0 \le t \le s-1$ is the maximum integer 
such that $i_{t+1}>i_t$.

Let us turn to another Bethe ansatz result, the character 
formula called combinatorial completeness of the string hypothesis 
at $q=0$ \cite{KN}:
\begin{align}
&\prod_{i=1}^L \hbox{ch}B^{r_i,l_i} 
= \sum_{m}\Omega(m)e^{\lambda(m)},\label{eq:comp}\\
&\Omega(m) = \det F \prod_{(a,j) \in H}
\frac{1}{m^{(a)}_j}
\binom{p^{(a)}_j+ m^{(a)}_j - 1}{m^{(a)}_j - 1} \quad \in \Z,
\label{omega}
\end{align}
where $\binom{s}{t} = s(s-1)\cdots(s-t+1)/t!$ and 
$\hbox{ch}B^{r,l}$ 
is the character of $B^{r,l}$.
$\lambda(m), \,p^{(a)}_j$ and $F$ are defined by 
(\ref{lamH}), (\ref{eq:vacancy}) and (\ref{F}).
The sum in (\ref{eq:comp}) extends over 
all $m^{(a)}_j \in \Z_{\ge 0}$ canceling out exactly leaving  
the character of $B$.
(\ref{eq:comp}) and (\ref{omega}) are the special cases of 
eq.(5.13) and eq.(4.1) in \cite{KN}, respectively.
$\Omega(m)$ (denoted by $R(\nu, N)$ therein) is the number of 
off-diagonal solutions to the string center equation with 
string content $m$.
It is known (\cite{KN} Lemma 3.7) that $\Omega(m) \in \Z_{\ge 1}$ 
provided that $p^{(a)}_j \ge 0$ for all $(a,j) \in H$.

\section{Dynamical period and state counting}\label{sec:4}

In (\ref{eq:pm}), the soliton content $m=(m^{(a)}_j)$ 
is introduced as the conserved quantity associated with the 
commuting transfer matrices.
One the other hand, the 
$m^{(a)}_j$ in the string content $m=(m^{(a)}_j)$ 
is the number of strings of color $a$ and 
length $j$ in the Bethe ansatz in section \ref{sec:3}.
{}From now on we identify them motivated 
by the factor $\zeta^{-E^{(r)}_l}$ in 
(\ref{la0}) and some investigation of Bethe vectors at $q=0$.
In view of Conjecture \ref{conj:adm}, the data of the form 
$m=(m^{(a)}_j)$ is defined to be a {\em content} if and only if 
$p^{(a)}_j \ge 0$ for all $(a,j) \in H$.
Thus $\det F > 0$ and 
$\lambda(m)$ in (\ref{lamH}) is a dominant weight for any content $m$.
\begin{conjecture}\label{conj:dp}
If $p \in P(m)$ and $(T^{(r)}_l)^t(p)\neq 0$ for any $t$, 
the dynamical period of $p$ under $T^{(r)}_l$ 
(minimum positive integer $t$ such that $(T^{(r)}_l)^t(p)=p$)
is equal to ${\mathcal P}^{(r)}_l$ $(\ref{lcm})$ generically and 
its divisor otherwise.
\end{conjecture}
In the situation under consideration, the whole $T^{(r)}_l$ orbit of $p$ 
belongs to $P(m)$ due to Proposition \ref{prop:exist}.
Naturally we expect $(\Lambda^{(r)}_l)^{\mathcal{P}^{(r)}_l} = 1$, which can 
indeed be verified for $A^{(1)}_1$.
Conjecture \ref{conj:dp}
has been checked, for example in $A^{(1)}_3$ case, 
for $B=(B^{1,1})^{\otimes 3}\otimes B^{2,2}$ and 
$B^{2,1}\otimes B^{2,1} \otimes B^{3,1} \otimes B^{3,2}$.

Let us present more evidence of Conjecture \ref{conj:dp}.
To save the space, $\cdot=\otimes$ is dropped 
when $B=(B^{1,1})^{\otimes L}$.
In each table, the period under 
$T^{(r)}_l$ with maximum $l$ is equal to that under $T^{(r)}_\infty$.
%


\vspace{0.3cm}\noindent
$A^{(1)}_1$, 
state $= 1221121122221$, 
\; content $= ((321))$

\def\arraystretch{1.2}

\begin{tabular}{c|cccc|c}
$(r,l)$ & \multicolumn{4}{|c|}{LCM of} & = period \\ \hline
(1,1) & 1,& 13, & 13,& 13 & 13 \\
(1,2) & 1,& $\frac{91}{3}$,  &$\frac{91}{16}$, &$\frac{91}{16}$&91 \\
(1,3) & 1,&91, &$\frac{273}{16}$,  &$\frac{273}{107}$& 273
\end{tabular}

\vspace{0.23cm}\noindent
$A^{(1)}_1$, 
state $= 122 \cdot 112 \cdot 12 \cdot 1222 \cdot 2 \cdot 11111 \cdot 
1122 \cdot 111$, 
\; content $= ((4321))$

\begin{tabular}{c|ccccc|c}
$(r,l)$ & \multicolumn{5}{|c|}{LCM of} &  = period \\ \hline
(1,1) & 1,& 2 &  &  &  &2 \\
(1,2) & 1,&7, &$\frac{7}{2}$, &21, &42 &42 \\
(1,3) & 1,& 14,  &7, &$\frac{21}{4}$,  &$\frac{21}{2}$&42\\
(1,4) & 1,&21, &$\frac{21}{2}$, &$\frac{63}{8}$, &$\frac{126}{29}$& 126
\end{tabular}


\vspace{0.23cm}\noindent
$A^{(1)}_3$,
 state $= 134 \cdot 34 \cdot 1 \cdot 134 \cdot 23 \cdot 1 \cdot 13$, 
\; content $= ((432),(31),(1))$

\begin{tabular}{c|ccccccc|c}
$(r,l)$ & \multicolumn{7}{|c|}{LCM of}  &  = period \\ \hline
(1,1) & 1,&$ \frac{380}{39}$,&$ \frac{95}{6}$,&$ \frac{95}{6}$,
&$ \frac{380}{31}$,&$ \frac{380}{27}$, &$\frac{380}{29}$&380  \\
(1,2) & 1,&  $\frac{190}{39}$, &$\frac{95}{12}$, &$\frac{95}{12}$, 
&$\frac{190}{31}$, &$\frac{190}{27}$, &$\frac{190}{29}$ & 190 \\
(2,1) & 1,& $\frac{190}{13}$, &$\frac{95}{4}$, &$\frac{95}{4}$, 
&$\frac{190}{137}$, &$\frac{190}{9}$,&$\frac{190}{73}$ & 190 \\
(2,2)& 1, & $\frac{76}{5}$, &$\frac{38}{3}$, &$\frac{38}{3}$,
 &$\frac{76}{41}$, &$\frac{76}{21}$, &$\frac{76}{31}$ & 76 \\
(2,3)&1, &$\frac{95}{6}$, &$\frac{95}{11}$, &$\frac{95}{11}$,
&$\frac{95}{34}$, &$\frac{95}{48}$,&$\frac{95}{41}$ & 95 \\
(3,1)&1, &$\frac{380}{13}$, &$\frac{95}{2}$, &$\frac{95}{2}$, 
&$\frac{380}{137}$, &$\frac{380}{9}$, &$\frac{380}{263}$& 380 
\end{tabular}
\[
\hspace{-31.5mm}
\mbox{$A^{(1)}_3$, state}=
233 \cdot 
\begin{array}{cc}
1 \!&\! 2 \\[-2mm]
2 \!&\! 3 \\[-2mm]
3 \!&\! 4 
\end{array}
\cdot 
\begin{array}{cc}
1 \!&\! 1 \\[-2mm]
3 \!&\! 4 
\end{array}
\cdot 1,
\; \mbox{ content $= ((3),(3),(2))$}\vspace{-3mm}
\]
\indent
\begin{tabular}{c|cccc|c}
$(r,l)$ & \multicolumn{4}{|c|}{LCM of} & = period \\ \hline
(1,1) & 1, &$\frac{11}{2}$, &11, &22 & 22 \\
(1,2) & 1, &$\frac{11}{4}$, &$\frac{11}{2}$, &11 & 11 \\
(1,3) & 1, &$\frac{11}{6}$, &$\frac{11}{3}$, &$\frac{22}{3}$ & 22 \\
(2,1) & 1, &11, &$\frac{33}{7}$, &$\frac{66}{7}$ & 66 \\
(2,2) & 1, &$\frac{11}{2}$, &$\frac{33}{14}$, &$\frac{33}{7}$ & 33 \\
(2,3) & 1, &$\frac{11}{3}$, &$\frac{11}{7}$, &$\frac{22}{7}$ & 22 \\
(3,2) & 1, &11, &$\frac{33}{7}$, &$\frac{33}{20}$ & 33 
\end{tabular}
\vspace{0.3cm}

Here, content=$((3111),(44),(2))$ for example means that 
$m^{(1)}_1\!=\!3, m^{(1)}_3\!=\!m^{(3)}_2\!=\!1,
m^{(2)}_4\!=\!2$ 
and  the other $m^{(a)}_j$'s are $0$.

Let us turn to another application of the Bethe ansatz results
(\ref{eq:comp}) and (\ref{omega}).
We introduce 
$
{\mathcal T}(P(m))=
\bigcup_{a=1}^n\bigcup_{j\ge 1}\{T^{(a)}_j(p)\mid p \in P(m)\},
$
which is the subset of $B$ consisting of all kinds of one step 
time evolutions of $P(m)$.
Under Conjecture \ref{conj:convergent}, any state 
$p \in P(m)$ is $(a,j)$-evolvable for $j$ sufficiently large.
Thus {}from Proposition \ref{prop:exist}, $p$ 
is expressed as $p = (T^{(a)}_j)^k(p)$ for some $k$, 
showing that ${\mathcal T}(P(m)) \supseteq P(m)$.
In general ${\mathcal T}(P(m))$ can contain 
non-evolvable states which do not belong to $P(m)$.
\begin{conjecture}\label{conj:omega}
For any content $m$ such that ${\mathcal T}(P(m))=P(m)$,
the following relation holds:
\begin{equation}\label{eq:omega}
\Omega(m) =\frac{\vert P(m)\vert}{\vert W\lambda(m)\vert}.
\end{equation}
\end{conjecture}
In view of Remark \ref{rem:inv} and 
Conjecture \ref{conj:adm}, the right hand side is the 
number of states in the periodic $A^{(1)}_n$ automaton
having the content $m$ and a fixed weight.
Thus it is equal to 
$\sharp\{p \in P(m) \mid {\rm wt} p=\lambda(m)\}$.
In case ${\mathcal T}(P(m))\supsetneqq P(m)$, we expect that 
$\vert P(m)\vert/\vert W\lambda(m)\vert$
is a divisor of $\Omega(m)$.

Let us present two examples of Conjecture \ref{conj:omega}.
In the periodic $A^{(1)}_3$ automaton with 
$B=B^{1,2}\otimes B^{1,1} \otimes B^{1,2} \otimes B^{1,1}$, 
there are 1600 states among which 824 are evolvable.
They are classified according to the contents $m$ in the following table.
\begin{center}
\begin{tabular}{c|c|c|c|c}
$m$ & $\lambda(m)$ & ${\vert W\lambda(m)\vert}$
& $\vert P(m)\vert$ & $\Omega(m) $ \\ \hline
$(\emptyset,\emptyset,\emptyset)$   & $(6,0,0,0)$ & 4    & 4  & 1 \\
$((1),\emptyset,\emptyset)$       & $(5,1,0,0)$ & {12} & 48 & 4 \\
$((11),\emptyset,\emptyset)$      & $(4,2,0,0)$ & {12} & 24 & 2 \\
$((2),\emptyset,\emptyset)$       & $(4,2,0,0)$ & {12} & 72 & 6 \\
$((21),\emptyset,\emptyset)$      & $(3,3,0,0)$ & {6}  & 24 & 4 \\
$((3),\emptyset,\emptyset)$       & $(3,3,0,0)$ & {6}  & 36 & 6 \\
$((11),(1),\emptyset)$       & $(4,1,1,0)$ & {12} &  96 &  8 \\
$((22),(2),\emptyset)^{*}$   & $(2,2,2,0)$ & {4} &  24 & 12 \\
$((21),(1),\emptyset)$       & $(3,2,1,0)$ &{24} & 432 & 18 \\
$((111),(11),(1))$ & $(3,1,1,1)$ & {4} & 16 & 4 \\
$((211),(11),(1))$ & $(2,2,1,1)$ & {6} & 48 & 8 
\end{tabular}
\end{center}
\noindent
In the second column, 
$(\lambda_1,\lambda_2,\lambda_3,\lambda_4)$ means 
$\lambda(m) = (\lambda_1-\lambda_2)\Lambda_1+
 (\lambda_2-\lambda_3)\Lambda_2+ (\lambda_3-\lambda_4)\Lambda_3$.
In the last two cases, the subsets of $P(m)$ having 
the dominant weight $\lambda(m)$ are given by
\begin{align*}
&
\{ 
11 \!\cdot\! 2 \!\cdot\! 13 \!\cdot\! 4,\;
12 \!\cdot\! 3 \!\cdot\! 14 \!\cdot\! 1,\;
13 \!\cdot\! 4 \!\cdot\! 11 \!\cdot\! 2,\;
14 \!\cdot\! 1 \!\cdot\! 12 \!\cdot\! 3\}
\;\;\hbox{for}\; m=((111),(11),(1)),\\
&
\{
11 \!\cdot\! 2 \!\cdot\! 23 \!\cdot\! 4,\;
12 \!\cdot\! 2 \!\cdot\! 13 \!\cdot\! 4, \;
12 \!\cdot\! 3 \!\cdot\! 24 \!\cdot\! 1, \;
13 \!\cdot\! 4 \!\cdot\! 12 \!\cdot\! 2,\\
&\;\; 14 \!\cdot\! 1 \!\cdot\! 22 \!\cdot\! 3,\;
22  \!\cdot\! 3 \!\cdot\! 14 \!\cdot\!1,\;
23 \!\cdot\! 4 \!\cdot\! 11 \!\cdot\! 2,\;
24 \!\cdot\! 1 \!\cdot\! 12 \!\cdot\! 3\}
\;\;\hbox{for}\; m=((211),(11),(1)).
\end{align*}
In the case of $B=B^{2,1}\otimes B^{2,1}\otimes B^{2,2}$, 
there are 720 states among which 518 are evolvable.
\begin{center}
\begin{tabular}{c|c|c|c|c}
$m$ & $\lambda(m)$ & ${\vert W\lambda(m)\vert}$
& $\vert P(m)\vert$ & $\Omega(m) $ \\ \hline
$(\emptyset,\emptyset,\emptyset)$   & $(4,4,0,0)$ & 6    & 6  & 1 \\
$(\emptyset,(1),\emptyset)$       & $(4,3,1,0)$ &  24  & 72 & 3 \\
$(\emptyset,(11),(1))$              & $(4,2,1,1)$ &  12  & 36 & 3 \\
$(\emptyset,(2),\emptyset)$       & $(4,2,2,0)$ &  12  & 48 & 4 \\
$((1),(11),\emptyset)$              & $(3,3,2,0)$ &  12  & 36 & 3 \\
$((1),(11),(1))$                      & $(3,3,1,1)$ &   6  & 72 & 12 \\
$((1),(21),(1))$                      & $(3,2,2,1)$ &  12  & 240 & 20 \\
$((2),(22),(2))^{*}$                  & $(2,2,2,2)$ &   1  &  8  & 32 \\
\end{tabular}
\end{center}
\noindent
The assumption ${\mathcal T}(P(m))=P(m)$ of the conjecture is
valid for all the contents except $((22),(2),\emptyset)$ and 
$((2),(22),(2))$ 
marked with $\ast$.

\end{document}